# Bonding Distances as Exact Sums of the Radii of the Constituent Atoms in Nanomaterials - Boron Nitride and Coronene

**Raji Heyrovska**

Institute of Biophysics, Academy of Sciences of the Czech Republic.

Email: rheyrovs@hotmail.com

**Abstract**

This paper presents for the first time the exact structures at the atomic level of two important nanomaterials, boron nitride and coronene. Both these compounds are hexagonal layer structures similar to graphene in two dimensions and to graphite in three-dimensions. However, they have very different properties: whereas graphene is a conductor, h-BN is an electrical insulator and coronene is a polycyclic aromatic hydrocarbon of cosmological interest. The atomic structures presented here are based on bond lengths as the sums of the atomic radii.

**Introduction**

An introduction to the properties of boron nitride and coronene can be found in [1, 2] and in the references to the original articles therein. Boron nitride is an inorganic compound and the hexagonal form consists of alternating boron and nitrogen atoms. The BN bond length in the regular hexagon is 0.145 nm (Fig. 1a), [3]. In the solids, the hexagonal layers are bound by van der Waals forces at a uniform distance of 0.33 nm as in graphite (Fig. 1b), [3]. BN has excellent insulating and lubricating properties and high thermal and chemical stability. It is readily machinable and finds extensive use in ceramics, high-temperature equipments, cosmetics, etc. [1, 4 - 6]. It also forms nanotubes [3] and nanomesh, both of which have many applications [1, 7 - 9].

Coronene is a highly symmetrical polycyclic aromatic hydrocarbon (PAH) consisting of a flat ring of six fused benzene molecules (Fig. 1c), [10]. Whereas, coronene is an aromatic hexagonal compound consisting of carbon and hydrogen atoms, graphene consists of hexagons of *only* carbon atoms. Crystals of coronene are parallel stacks of layers of the PAH held at the inter-layer distances of about 0.35 nm [11] by van der Waals forces, as in graphite. Coronene molecules are also found in the interstellar medium, in comets, and in meteorites and are potential indicators of organic life elsewhere in the universe [10-13]. The molecular structure and vibrational motion of aromatic hydrocarbons have posed considerable theoretical challenges.

## The atomic structure of h-boron nitride and coronene

Here, the author's previous work on the atomic structures of hexagonal benzene and graphene [14] has been extended to the above compounds, h-BN and coronene. The known bond lengths in both cases have been divided here into the radii of the constituent atoms. Thereby, the precise structures of these nanomaterials have been established at the atomic level. The BN bond length d(BN) = 0.145 nm [3] and the single bond radius of $R(N_{s.b.}) = 0.070$ nm [15]. The radius of boron, obtained as the difference $R(B_{res}) = 0.145 - 0.070 = 0.075$ nm (subscripts s.b.: single bond, res.: resonance bond). This value of the radius of B is about midway between those of the single bond (0.081 nm) and double bond (0.071 nm) and hence can be considered as the resonance bond radius $R(B_{res})$, similar to the resonance bond radius, $R(C_{res})$ of carbon in graphene [14]. Thus the exact atomic structure of h-BN could be established as shown in Fig. 2.

In the case of coronene, C atoms of covalent double bond radius, $R(C_{d.b.}) = 0.067$ nm (subscript d.b.: double bond) alternate with those having resonance bond radius, $R(C_{res}) = R(C_{gr}) = 0.071$ nm (subscript gr.: graphite) and the bond length $d(C_{d.b.}C_{r.b.}) = d(C_{d.b.}) + d(C_{res}) = 0.067 + 0.071 = 0.138$ nm as for benzene [14]. See Fig. 3 for the precise structure of coronene, where the covalent radius of H is $R(H) = 0.037$ nm, and the CH bond lengths, $d(C_{res}H) = 0.108$ nm, $d(C_{d.b.}H) = 0.104$ nm.

Fig. 4 for graphene (see also [14], [16]) is given for comparison, where all C atoms have the same radii $R(C_{res}) = 0.071$ nm. The bond length, $d(C_{res}C_{res}) = 2R(C_{res}) = R(C+)_{res} + R(C-)_{res}$ is the sum of the Golden ratio ($\varphi = 1.618..$) based cationic and anionic radii as described in [14b]. These alternating positively and negatively charged ionic resonance forms are probably responsible for the electrical conductivity of graphene. The ionic forms are shown here in Fig. 4 as red and blue circles of radii $R(C+)_{res} = d(C_{res}C_{res})/\varphi^2 = 0.054$ nm and $R(C-)_{res} = d(C_{res}C_{res})/\varphi = 0.088$ nm respectively. Figs. 2 - 4 have been drawn to scale.

Thus, the author has presented here, for the first time, the ultimate atomic structures of two hexagonal nanomaterials, h-BN and coronene, and compared it with that of graphene on the basis of exact atomic radii which make up the bonding distances.

**Acknowledgements:** The author is grateful to the Institute of Biophysics of the Academy of Sciences of the Czech Republic (ASCR) for financial support.

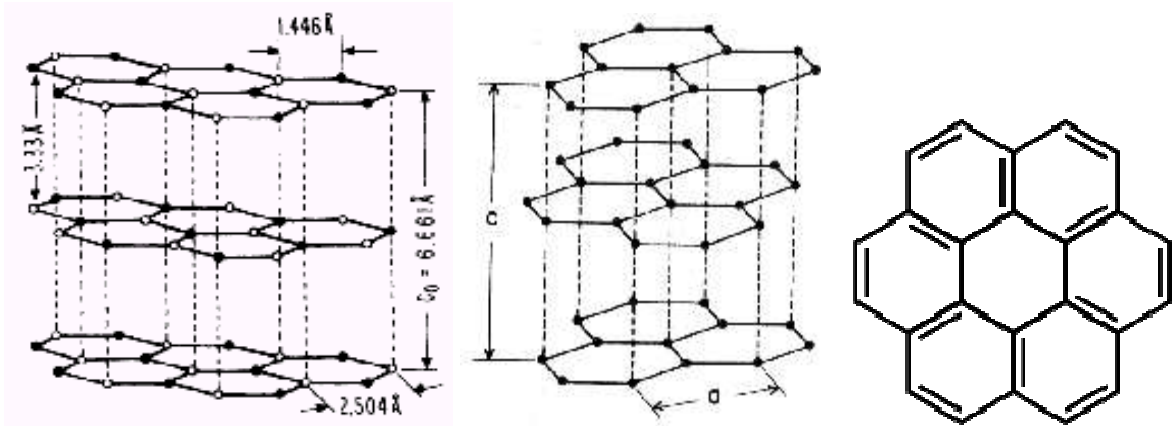

**Fig. 1 a-c:** Left to right: boron nitride [3], graphite (a = 0.246, c = 0.670 nm) [3], coronene [10].

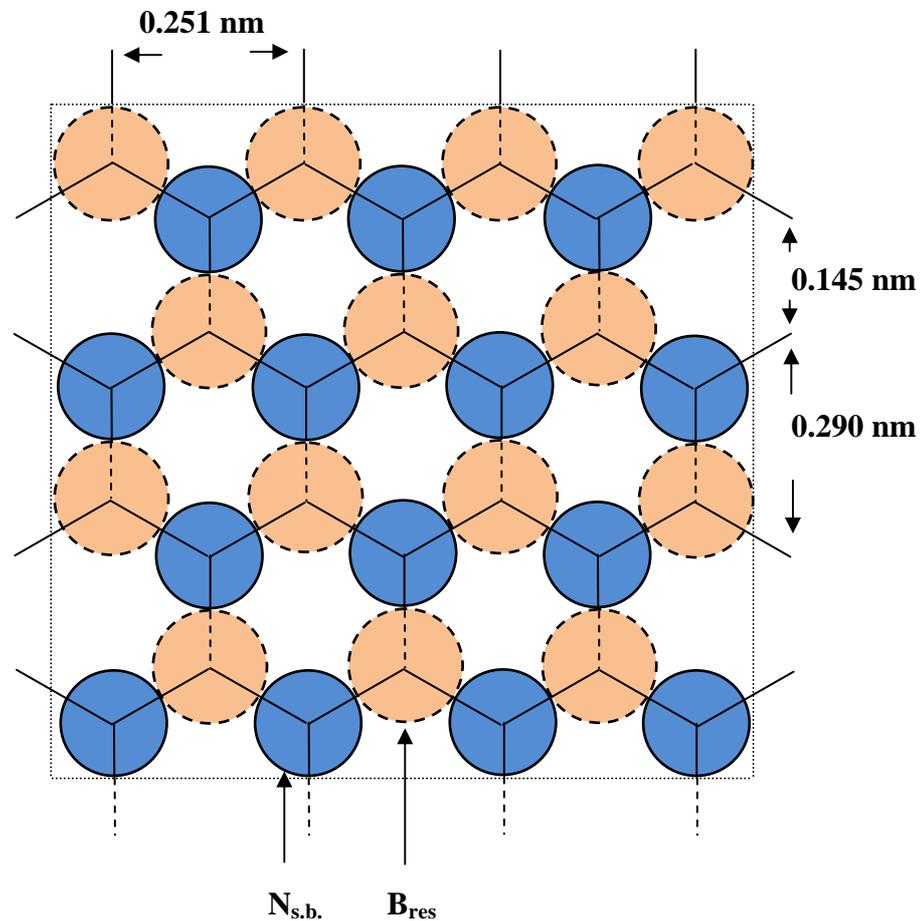

**Fig. 2**. Atomic structure of hexagonal boron nitride (h-BN). B (copper color), N (blue). Bond length, d(BN) = 0.145 nm = R(B$_{res.}$) + R(N$_{s.b.}$). Area of the rectangle containing 14 atoms each of B and N = 0.90 x 0.87 = 0.78 nm$^2$.

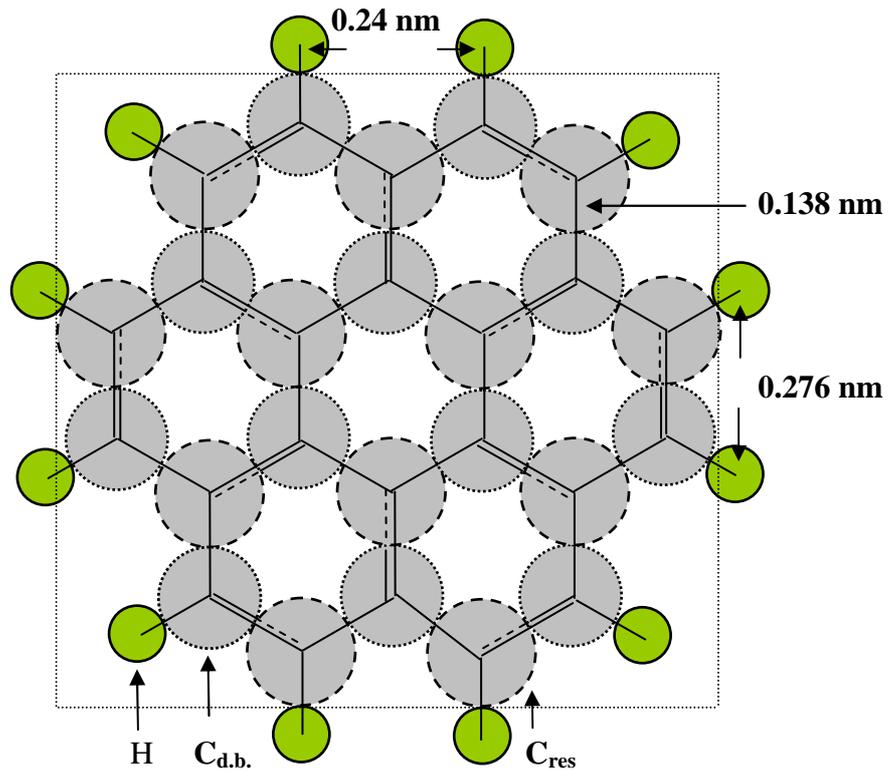

**Fig. 3** (above). Atomic structure of coronene. Bond length, $d(C_{d.b.}C_{res.}) = 0.138$ nm as in benzene [14]. Area of the rectangle containing the 24 atoms of C = 0.86 x 0.83 = 0.71 nm².

**Fig. 4** (below). Atomic structure of graphene. Bond length $d(CC)_{res} = 2R(C)_{res} = 0.142$ nm. The alternate red and blue circles are cations and anions of radii sum, $d(CC)_{res} = R(C+)_{res} + R(C-)_{res} = 0.142$ nm. Area of the rectangle containing 28 carbon atoms = 0.88 x 0.85 = 0.75 nm².

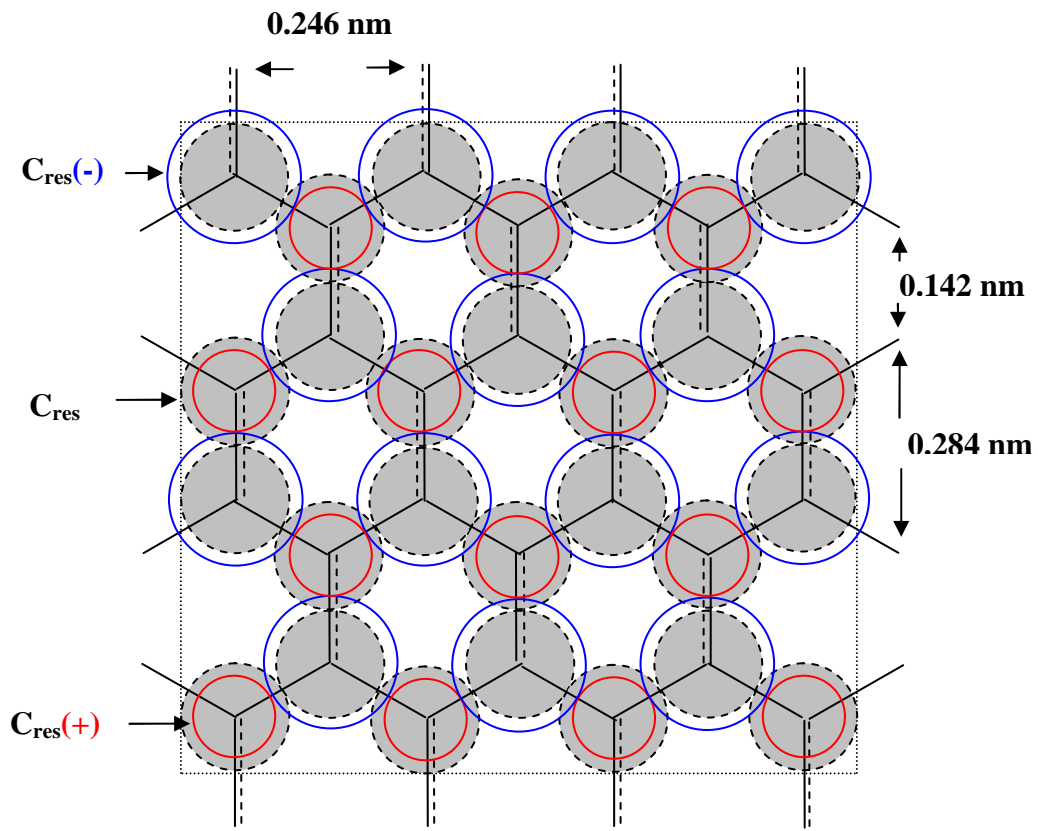